\documentclass[showpacs,twocolumn,preprintnumbers,amsmath,amssymb]{revtex4}
\usepackage{graphicx}
\usepackage{subfigure}
\usepackage{dcolumn}
\usepackage{bm}
\begin{document}
\title{Operational typicality of the nonequilibrium states:\\
 a thermodynamic lower bound of the deviation from equilibrium}
\author{Takaaki Monnai}%
\email{monnai@suou.waseda.jp}%
\affiliation{$*$Department of Applied Physics, Osaka City University,
3-3-138 Sugimoto, Sumiyoshi-ku, Osaka 558-8585, Japan}
\begin{abstract}
The typicality of the canonical state shows that majority of the states are indistinguishable from  equilibrium, and thus the nonequilibrium states are exceptionally rare in the extremely high-dimensional Hilbert space.  
On the contrary, we can easily apply an external force acting on the system, and then the actual density matrix quantitatively deviates from the canonical state specified by the system Hamiltonian at each instance.  
To express how the external forcing amounts to the deviation from equilibrium,  we give a universal thermodynamic expression of the lower bound of Hilbert-Schmidt distance between the actual nonequilibrium and corresponding canonical states. The lower bound is expressed only by  the amount of forcing and its consequent entropy production rate.
\end{abstract}    
\maketitle
There are two apparent contradicting statements on the non-typicality of the states out of equilibrium.  
First, for macroscopic quantum systems consisting of a system of interest $S$ and a large reservoir $R$, the system state $\hat{\rho}_s$ is typically indistinguishable from canonical equilibrium state $\hat{\rho}_{eq}$ in the Hilbert space from informational point of view\cite{Lebowitz,Popesucu}. This canonical typicality means that the  nonequilibrium states are considered as extremely rare compared with the equilibrium states. 
On the contrary, we can easily apply various external forcing to the system $S$ such as manipulation of latex by optical tweezers, fine tuning of gate voltage in quantum dots, and any other thermodynamic operations.
And the system state is then driven out of equilibrium. Therefore in this operational way, we can systematically prepare nonequilibrium states with a considerable deviation from equilibrium.
It is thus questionable to regard the nonequilibrium states as exceptional.

As a first step to rationalize the paradox that the nonequilibrium states are rare but easily prepared, we need a {\it lower bound of the deviation} from equilibrium.
This is in marked contrast to the typicality of canonical states\cite{Lebowitz,Popesucu}, which essentially provide upper bounds of the deviation from equilibrium based on the probabilistic arguments.   
Here it is remarked that the quantum ergodic theorems\cite{Lebowitz2,vonNeumann} discuss on the expectation values of only a limited class of macroscopic quantitites\cite{vanKampen}, which do not contain the full information of the actual state or the density matrix, and in 
these cases preparation of initial states with considerable deviation of the quantities of interest is actually difficult.
We emphasize that our quantity of interest is the density matrix itself.  

In this letter, we thermodynamically express the lower bound of this deviation  $\hat{\rho}_s(t)-\hat{\rho}_{eq}(t)$ between the actual nonequilibrium state of the system $\hat{\rho}_s(t)$ and the equilibrium $\hat{\rho}_{eq}(t)$ specified by the instantaneous system Hamiltonian $\hat{H}_s(t)$ with the use of a strength of the external forcing and its consequent entropy production. 

\section*{}     
We consider a time evolution of an externally forced system which weakly interacts with a large reservoir\cite{Remark1}. The evolution of the total system(system and reservoir) is unitary.
We externally manipulate the system during the initial and the final time $0\leq t\leq {\cal T}$. The reservoir is assumed to be large enough so that its thermodynamic properties such as temperature are regarded as constant during our operation.   
The total Hamiltonian 
\begin{equation}
\hat{H}(t)=\hat{H}_s(t)+\hat{H}_r+\hat{V}
\end{equation}
is the sum of those of the system $\hat{H}_s(t)$, the reservoir $\hat{H}_r$, and the interaction between them $\hat{V}$, respectively.
Initially, the system and the reservoir are in an equilibrium at an inverse temperature $\beta=\frac{1}{k_BT}$. Here $k_B$ is the Boltzmann constant, and $T$ is the absolute temperature.	
The system Hamiltonian $\hat{H}_s(t)$ is time dependent due to the external forcing during $0\leq t\leq {\cal T}$. Then the system state becomes out of equilibrium.
From microscopic point of view, the initial total density matrix $\hat{\rho}(0)$ evolves as $\hat{U}(t)\hat{\rho}(0)\hat{U}(t)^+$ by a unitary operator $\hat{U}(t)\equiv{\rm T}\{e^{-\frac{i}{\hbar}\int_0^t\hat{H}(s)ds}\}$, which is given as a solution of the Schr\"odinger equation 
$i\hbar\frac{\partial \hat{U}(t)}{\partial t}=\hat{H}(t)\hat{U}(t)$
with the initial condition $\hat{U}(0)=\hat{1}$.
And we shall evaluate how the actual reduced density matrix 
\begin{equation}
\hat{\rho}_s(t)\equiv{\rm Tr}_r\hat{U}(t)\hat{\rho}(0)\hat{U}(t)^+
\end{equation}
deviates from the instantaneous canonical state
\begin{equation}
\hat{\rho}_{eq}(t)\equiv\frac{1}{Z_s(t)}e^{-\beta \hat{H}_s(t)}, \label{canonicalstate}
\end{equation}
which is specified by the system Hamiltonian $\hat{H}_s(t)$ and the inverse temperature of the reservoir $\beta$. Here the partition function is defined from the normalization $Z_s(t)\equiv{\rm Tr}_se^{-\beta \hat{H}_s(t)}$.
\vspace{1cm}
In order to evaluate the deviation from equilibrium, we first note that the dissipative work $W_{diss}$ measures how far the system is deviating from equilibrium\cite{Esposito1,Callen} which is well-defined even far out of equilibrium\cite{Esposito1,Suzuki}.
The dissipative work  
\begin{eqnarray}
W_{diss}\equiv W-\Delta F
\end{eqnarray}
is the extra work required to achieve the nonequilibrium process aside from the minimum work given by the free energy change $\Delta F$ for the adiabatically slow reversible process.
Here the free energy change is defined from the partition functions of the system at initial and final times as $\Delta F\equiv-\frac{1}{\beta}\log\frac{Z_s(t)}{Z_s(0)}$.
The work is defined as the energy change due to the time dependence of the Hamiltonian\cite{Sekimoto,Jarzynski,Crooks} 
\begin{equation}
W\equiv\int_0^{\cal T} dt{\rm Tr}_s\{\hat{\rho}_s(t)\dot{\hat{H}}_s(t)\}, \label{work}
\end{equation}
where the reservoir variables are traced out, and the reduced density matrix $\
\hat{\rho}_s(t)={\rm Tr}_r\hat{U}(t)\rho\hat{U}(t)^+$ in the Hilbert space of the system states ${\cal H}_s$ gives the actual state, which deviates from the equilibrium state $\hat{\rho}_{eq}(t)=\frac{1}{Z_s(t)}e^{-\beta \hat{H}_s(t)}$. 
We don't need the explicit expression of the heat, however, we show the definition for completeness. 
From the first law of thermodynamics, the heat flowing to the system is automatically determined as the energy variation due to the change of the state\cite{Sekimoto}
\begin{equation}
Q\equiv\int_0^{\cal T} dt{\rm Tr}\{\frac{1}{i\hbar}[\hat{H}(t),\hat{U}(t)\hat{\rho}(0)\hat{U}^+(t)]\hat{H}_s(t)\}.
\end{equation}
Indeed, by taking the time derivative of the expectation value of the system Hamiltonian, the total energy change 
\begin{eqnarray}
&&{\rm Tr}\left(\{\hat{U}({\cal T})\hat{\rho}(0)\hat{U}^+({\cal T})\hat{H}_s({\cal T})\}-\hat{\rho}(0)\hat{H}_s(0)\right) \nonumber \\
&=&W+Q\label{work1}
\end{eqnarray}
is composed of the work done and the heat flowing to the system. 
We also note that the work done along the instantaneous canonical state given in Eq.(\ref{canonicalstate}) is equal to the free energy change 
\begin{eqnarray}
&&\int_0^{\cal T} dt {\rm Tr}_s\frac{1}{Z_s(t)}e^{-\beta \hat{H}_s(t)}\dot{\hat{H}}_s(t)  \nonumber \\
&=&\int_0^{\cal T} dt{\rm Tr}_s\frac{d}{dt}\left(\frac{-1}{\beta Z_s(t)}e^{-\beta \hat{H}_s(t)}\right) \nonumber \\
&&-\int_0^{\cal T} dt{\rm Tr}_s\frac{1}{\beta}\frac{\dot{Z}_s(t)}{Z_s(t)^2}e^{-\beta \hat{H}_s(t)} \nonumber \\
&=&-\int_0^t dt\frac{1}{\beta}\frac{\dot{Z}_s(t)}{Z_s(t)} \nonumber \\
&=&\Delta F.\label{free}
\end{eqnarray}
The first equality is from the integration by parts.
In the second equality, we used the normalization ${\rm Tr}_s\frac{1}{Z_s(t)}e^{-\beta H_s(t)}=1$ for all $t$ in the first and the second terms.  
The dissipative work is then calculated with the use of Eqs.(\ref{work},\ref{free}) as
\begin{eqnarray}
&&W_{diss} \nonumber \\
&=&W-\Delta F \nonumber \\
&=&\int_0^{\cal T} dt{\rm Tr}_s\{\left(\hat{\rho}_s(t)-\hat{\rho}_{eq}(t)\right)\dot{\hat{H}}_s(t)\} \nonumber \\
&\geq&0.\label{ineq}
\end{eqnarray}
The positivity is from the second law of thermodynamics.
The dissipative work expresses the entropy production in a nonequilibrium process of an isothermal system.  
This equation (\ref{ineq}) is well-known as the minimum work principle in thermodynamics\cite{Jarzynski}, however, it is remarkable that the distance $\hat{\rho}_s(t)-\hat{\rho}_{eq}(t)$ between the actual and equilibrium states explicitly appears. 

Then our first result is directly obtained from Eq.(\ref{ineq}).       
Let us apply the Cauchy-Schwartz inequality\cite{Thirring} to the integrand as 
\begin{eqnarray} 
&&{\rm Tr}_s\{\left(\hat{\rho}_s(t)-\hat{\rho}_{eq}(t)\right)\dot{\hat{H}}_s(t)\} \nonumber \\
&\leq&\left({\rm Tr}_s(\hat{\rho}_s(t)-\hat{\rho}_{eq}(t))^2\right)^{\frac{1}{2}}\left({\rm Tr}_s\dot{\hat{H}}_s(t)^2\right)^{\frac{1}{2}},\label{ineq2}
\end{eqnarray}
which disentangles the deviation from equilibrium $\hat{\rho}_s(t)-\hat{\rho}_{eq}(t)$ and the external forcing $\dot{\hat{H}}_s(t)$.
Then from Eqs.(\ref{ineq},\ref{ineq2}), we have a model-independent lower bound of the Hilbert-Schmidt distance between the actual and the canonical states 
\begin{equation}
\left({\rm Tr}_s(\hat{\rho}_s(t)-\hat{\rho}_{eq}(t))^2\right)^{\frac{1}{2}}\geq\frac{|\dot{W}_{diss}(t)|}{\left({\rm Tr}_s \dot{\hat{H}}_s(t)^2\right)^{\frac{1}{2}}}. \label{distance}
\end{equation}
Here the lower-bound solely depends on the external perturbation $\dot{\hat{H}}_s(t)$ and its consequent dissipative work done per unit time $\dot{W}_{diss}(t)$.
This relation directly shows that the external forcing yields dissipative work, and simultaneously the state deviates from the equilibrium.  
Especially, in the perturbation regime of weak forcing, we have the linearity of $\frac{|\dot{W}_{diss}(t)|}{\left({\rm Tr}_s\dot{\hat{H}}_s(t)^2\right)^{\frac{1}{2}}}$ averaged over a short time with respect to the forcing $\left({\rm Tr}_s\dot{\hat{H}}_s(t)^2\right)^{\frac{1}{2}}$ as shown in Fig.2.
Then, the right hand side of Eq.(\ref{distance}) is proportional to the external forcing $\left({\rm Tr}_s\dot{\hat{H}}_s(t)^2\right)^{\frac{1}{2}}$ at the lowest order, which explicitly shows that more stronger the forcing is, the more the actual state deviates from equilibrium. 
Namely, in the linear regime the distance is expressed as 
\begin{equation}
\left({\rm Tr}_s(\hat{\rho}_s(t)-\hat{\rho}_{eq}(t))^2\right)^{\frac{1}{2}}\propto \left({\rm Tr}_s\dot{\hat{H}}_s(t)^2\right)^{\frac{1}{2}},
\end{equation}
which shows that the deviation is of the first order with respect to the forcing $\left({\rm Tr}\dot{\hat{H}}_s(t)^2\right)^{\frac{1}{2}}$.
\section*{}
In Fig.1, we have numerically demonstrated how the distance $\left({\rm Tr}_s(\hat{\rho}_s(t)-\hat{\rho}_{eq}(t))^2)\right)^{\frac{1}{2}}$ is bounded by the thermodynamic quantity $\frac{|\dot{W}_{diss}(t)|}{\left({\rm Tr}_s\dot{\hat{H}}_s(t)^2\right)^{\frac{1}{2}}}$ for a $N$-site quantum spin chain in a spatially inhomogeneous oscillating magnetic field\cite{Monnai1}.
We consider the nearest neighbor exchange coupling and on-site magnetic energies.
More precisely, the total Hamiltonian $H(t)$ is given as
\begin{eqnarray}
&&\hat{H}(t)=\hat{H}_s(t)+\hat{H}_r+\hat{V} ;\nonumber \\
&&
\begin{cases}
\hat{H}_s(t)\equiv\sum_{j=1}^{N_s}\left(-J\hat{\sigma}_j^z\hat{\sigma}_{j+1}^z+\hat{\sigma}_j^x+h(t)\hat{\sigma}_j^z\right) & \\
\hat{H}_r\equiv\sum_{j=N_s+2}^N\left(-J\hat{\sigma}_j^z\hat{\sigma}_{j+1}^z+\hat{\sigma}_j^x+\hat{\sigma}_j^z\right) & \\
\hat{V}\equiv\epsilon\left(-J\hat{\sigma}_{N_s+1}^z\hat{\sigma}_{N_s+2}^z+\hat{\sigma}_{N_s+1}^x+\hat{\sigma}_{N_s+1}^z\right)&,
\end{cases}
\end{eqnarray}
where an oscillating magnetic field ${\bf B}(j,t)=(1,0,h(j,t))$ is externally applied with $h(j,t)=h_0\sin 2\alpha\pi t$ for $1\leq j\leq N_s$ and $h(j,t)=h_0$ for $j\geq N_s+1$.       
Namely, we regard the sites $j=1,...,N_s$ as the subsystem, $N_s+2\leq j\leq N$ as the reservoir, and $j=N_s+1$ as the interaction between them.
We choose the system size $N_s=2$, and the total system size $N=8$. We consider the ferromagnetic coupling $J=1$, edge of the linear regime $h_0=1$, and the weak coupling parameter is chosen as $\epsilon=0.1$\cite{Ramark1}.
Note that the thermodynamic properties can be verified for the relatively small system size\cite{Monnai1,Shankar}. This is because the $O(e^N)$ dimension of the Hilbert space does matter instead of the total system size $N$ itself.
There is a numerical support of this statement based on the system size dependence of the many body Hamiltonians for $6\leq N\leq 12$\cite{Monnai2}. 

Let us investigate the time evolution.
Initially, the system and the reservoir are in thermal equilibrium at an inverse temperature $\beta=1$, $\rho_s(0)=\frac{1}{Z_s(0)}e^{-\beta H_s(0)}$ and $\rho_r=\frac{1}{Z_r}e^{-\beta H_r}$.
Then we have numerically solved the unitary evolution by using a discretized evolution operator 
$\hat{U}(n\Delta t)=e^{-i\hat{H}(n\Delta t)\Delta t}e^{-i\hat{H}(n\Delta t)\Delta t}\cdot\cdot\cdot e^{-i\hat{H}(\Delta t)\Delta t}e^{-i\hat{H}(0)\Delta t}$ with a time step $\Delta t=0.02$ which is much shorter than the period of the oscillation.
Actually, compared with the numerical data for $\Delta t=0.05$, the results are regarded as well-converged.       
Importantly, in Fig.1, the lower bound $\left({\rm Tr}_s\dot{\hat{H}}_s(t)^2\right)^{\frac{1}{2}}$ is of the same order as the distance $\left({\rm Tr}_s(\hat{\rho}_s(t)-\hat{\rho}_{eq}(t))^2\right)^{\frac{1}{2}}$ for most time $t$.
This shows that we can roughly estimate the distance by calculating the ratio $\frac{|\dot{W}_{diss}(t)|}{\left({\rm Tr}_s\dot{\hat{H}}_s(t)^2\right)^{\frac{1}{2}}}$.  
The frequency of the oscillating magnetic field is chosen as $\alpha=0.2$ for Fig.1(a) and $\alpha=1$ for Fig.1(b), respectively. We have confirmed that qualitatively similar plots are obtained for other frequencies $\alpha\geq 0$. 
\begin{figure}
\center
{
\includegraphics[scale=0.8]{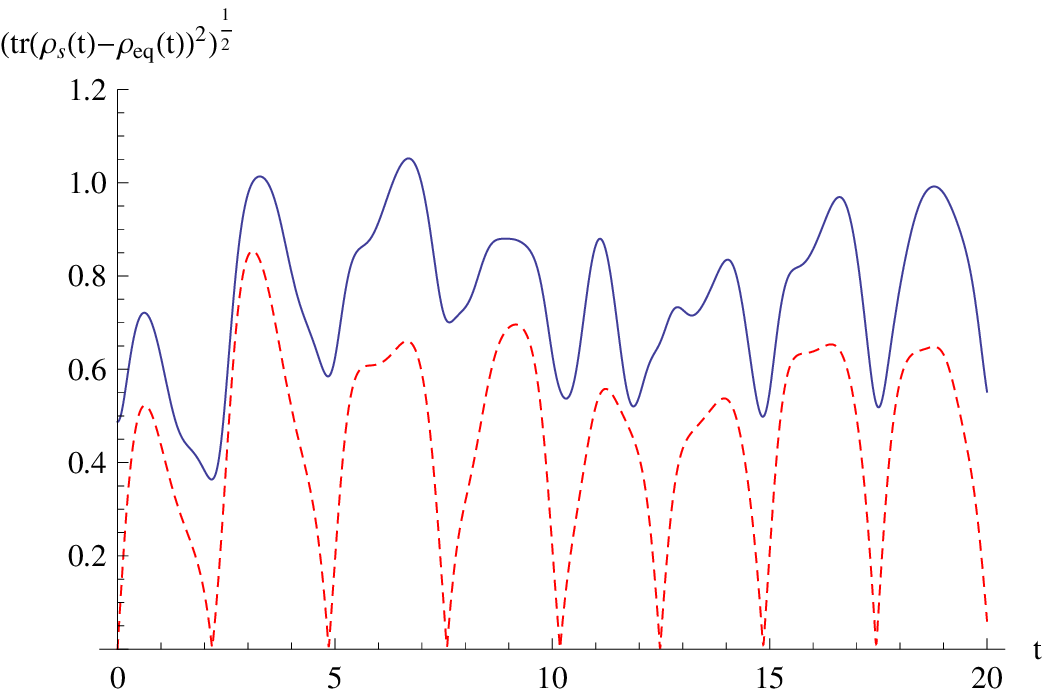}  
\includegraphics[scale=0.8]{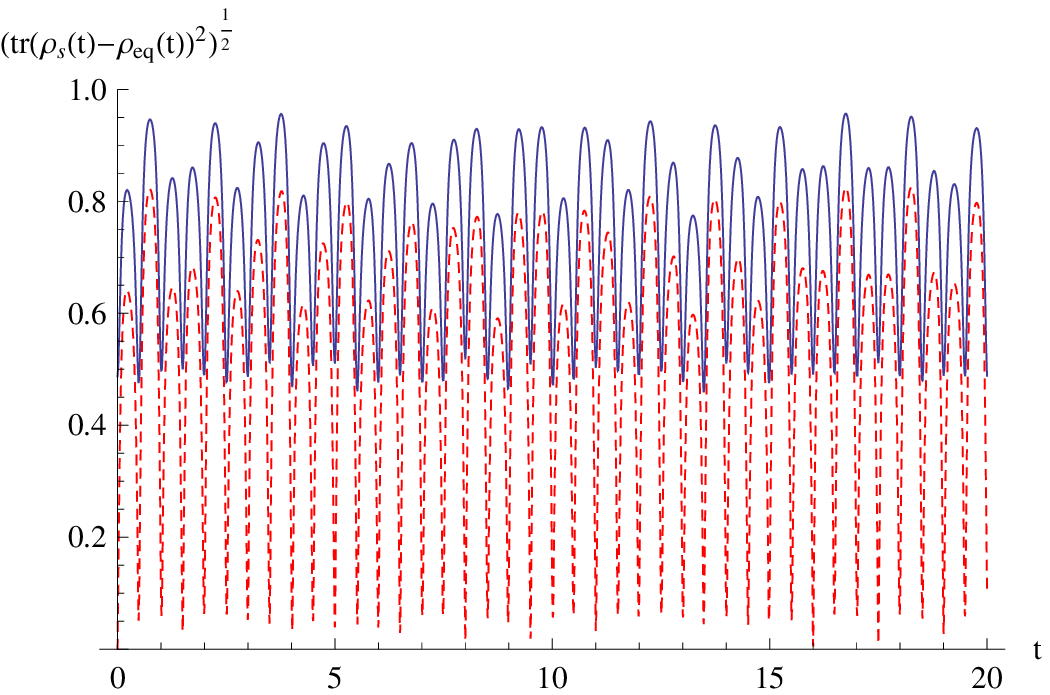}
}
\caption{Comparison between the distance $\left({\rm Tr}_s(\hat{\rho}_s(t)-\hat{\rho}_{eq}(t))^2\right)^{\frac{1}{2}}$(blue line) and the thermodynamic lower bound $\frac{|\dot{W}_{diss}|}{\left({\rm Tr}_s\dot{\hat{H}}_s(t)^2\right)^{\frac{1}{2}}}$(red dashed line). The distances temporally oscillate with the same frequency as the external magnetic field both for (a)$\alpha=0.2$ and (b)$\alpha=1$. The distance is small but not zero at $t=0$ due to the finite coupling strength.}
\end{figure} 
 
The lower bound in Eq.(\ref{distance}) $\frac{|\dot{W}_{diss}(t)|}{\left({\rm Tr}_s\dot{\hat{H}}_s(t)^2\right)^{\frac{1}{2}}}$ averaged over a short time $0.9\leq t\leq 1.1$ is shown in Fig.2 as a function of external forcing $\left({\rm Tr}_s\dot{\hat{H}}_s(t)^2\right)^{\frac{1}{2}}$ averaged over the same period $0.9\leq t\leq 1.1$.
We only changed the strength of the magnetic field $h_0$ from $0$ to $1$, and fixed the frequency $\alpha=0.2$.  
Then we observe that the distance linearly dependends on the external forcing.     
\begin{figure}
\center{
\includegraphics[scale=0.8]{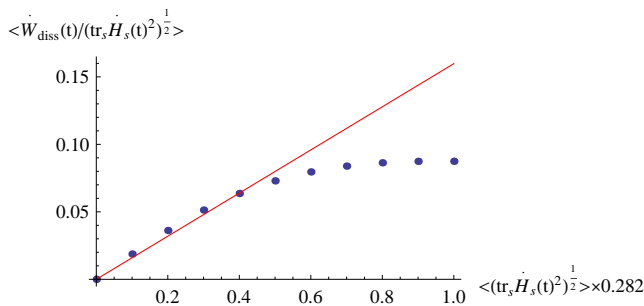}
}
\caption{Thermodynamic lower bound $\langle\frac{\dot{W}_{diss}(t)}{\left({\rm Tr}_s\dot{\hat{H}}_s(t)^2\right)^{\frac{1}{2}}}\rangle$ is plotted as a function of the external forcing $\langle\left({\rm Tr}_s\dot{\hat{H}}_s(t)^2\right)^{\frac{1}{2}}\rangle$. $\langle\rangle$ shows the average over a short time $0.9\leq t\leq 1.1$. The red line shows a numerical fitting.}
\end{figure}   
\section*{} 
As a measure of distance between the states, there are other choices such as the spectral norm.   
This is achieved by making the operators $\hat{\rho}_s(t)-\hat{\rho}_{eq}(t)$ and $\dot{\hat{H}}_s(t)$ positive, and apply the H\"older-inequality.
The positive operator $|\hat{A}|$ is defined for each observable $\hat{A}=\sum_{n}|A_n\rangle A_n\langle A_n|$ based on the spectral decomposition as
\begin{equation}
|\hat{A}|=\sum_{n}|A_n\rangle|A_n|\langle A_n|,
\end{equation}
where $|A_n|$ is the absolute value of the eigenvalue $A_n$.
For simplicity, we use the notation for the discrete spectrum, but the continuous case is also available.   
Then the H\"older inequality shows for $p,q\geq 0$ satisfying $\frac{1}{p}+\frac{1}{q}=1$, 
\begin{eqnarray}
&&\left({\rm Tr}_s|\hat{\rho}_s(t)-\hat{\rho}_{eq}(t)|^p\right)^{\frac{1}{p}}\left({\rm Tr}_s|\dot{\hat{H}}_s(t)|^q\right)^{\frac{1}{q}}\nonumber \\
&\geq&{\rm Tr}_s(\hat{\rho}_s(t)-\hat{\rho}_{eq}(t))\dot{\hat{H}}_s(t).
\end{eqnarray} 
Then by taking the limit $p\rightarrow\infty$, we have a lower bound for the distance measured by the spectral norm $\|\hat{A}\|\equiv{\rm sup}_{|\Psi\rangle\in {\cal H}_s}\frac{\langle\Psi|\; |\hat{A}|\; |\Psi\rangle}{\langle\Psi|\Psi\rangle}$ defined on the Hilbert space of the system ${\cal H}_s$ as
\begin{eqnarray}
&&\|\hat{\rho}_s(t)-\hat{\rho}_{eq}(t)\| \nonumber \\
&\geq&\frac{|\dot{W}_{diss}|}{{\rm Tr}_s|\dot{H}_s(t)|}.\label{distance2}
\end{eqnarray}
Eqs.(\ref{distance},\ref{distance2}) express the lower bound of the distance between the actual and canonical states by the dissipative work per unit time and the amount of forcing.
Note that another choice of the parameter $p$ and $q$ also gives measure of distance.
These thermodynamic expressions of the deviation from equilibrium are our main result.

Indeed, the presence of the lower-bound is in marked contrast with the typicality of the canonical state which essentially gives a typical upper bound, i.e. if we randomly choose a pure state $|\Psi\rangle$ from the Hilbert space ${\cal H}_s$ at an energy scale $E$ and weak coupling is satisfied, then the reduced density matrix ${\rm Tr}_r|\Psi\rangle\langle\Psi|$ is indistinguishable from the canonical equilibrium state with a probability very close to unity\cite{Lebowitz,Popesucu,Remark2}.

In conclusion, we derived universal lower bounds for the deviation from instantaneous equilibrium state due to the external forcing and its consequent dissipative work or entropy production.
It is remarkable that the lower bound does not explicitly depends on the model, and is thus regarded as thermodynamically universal.
The argument in this letter holds both for systems with finite and infinite dimensional Hilbert space, which is advantageous compared to the quantum ergodic theory.
\section*{}
This work is financially supported by JSPS research program under the Grant 22$\cdot$7744.


\begin{thebibliography}{30}
\bibitem{Remark1}The coupling is weak enough so that the reduced density matrix is described by a canonical ensemble when the total system is in thermal equilibrium.
\bibitem{Remark2}The probability to have a nonequilibrium state goes to zero in the macroscopic limit.
\end{thebibliography}
\end{document}